\documentclass[conference]{IEEEtran}
\IEEEoverridecommandlockouts

\usepackage{cite}
\usepackage{amsmath,amssymb,amsfonts}
\usepackage{graphicx}
\def\BibTeX{{\rm B\kern-.05em{\sc i\kern-.025em b}\kern-.08em
    T\kern-.1667em\lower.7ex\hbox{E}\kern-.125emX}}

\usepackage{bm}
\usepackage{blkarray}
\usepackage{dsfont}
\usepackage[english]{babel}
\usepackage{ifthen}
\usepackage{ifpdf}
\usepackage{hyperref}
\usepackage[nameinlink]{cleveref}
\usepackage{orcidlink}

\DeclareMathOperator{\trace}{Tr}

\begin{document}
\title{
Continuous Variable Quantum Key Distribution in Multiple-Input
Multiple-Output Settings
\thanks{This research is partly funded by the
School of Electronic and Electrical Engineering at the University of Leeds, and by Cisco Research via Silicon Valley Community Foundation.}
}

\author{\IEEEauthorblockN{\textsuperscript{*}Shradhanjali Sahu}
\IEEEauthorblockA{\textit{School of  Electronic and Electrical} \\
\textit{ Engineering, University of Leeds}\\
Leeds, UK \\
shradhanjali00sahu@gmail.com
\orcidlink{0000-0003-0447-2550}}
\and
\IEEEauthorblockN{Ahmed Lawey}
\IEEEauthorblockA{\textit{School of  Electronic and Electrical} \\
\textit{ Engineering, University of Leeds}\\
Leeds, UK \\
A.Q.Lawey@leeds.ac.uk}
\and
\IEEEauthorblockN{Mohsen Razavi}
\IEEEauthorblockA{\textit{School of  Electronic and Electrical} \\
\textit{ Engineering, University of Leeds}\\
Leeds, UK \\
M.Razavi@leeds.ac.uk \orcidlink{0000-0003-4172-2125}}}

\maketitle

\begin{abstract}
We investigate  quantum key distribution (QKD) in optical multiple-input-multiple-output (MIMO) settings. Such settings  can prove useful in dealing with harsh channel conditions as in, e.g., satellite-based QKD. We study a $2\times2$ setting for continuous variable (CV) QKD with Gaussian encoding and heterodyne detection and reverse reconciliation. We present our key rate analysis for this system and compare it with single-mode and multiplexed CV QKD scenarios. We show that we can achieve multiplexing gain using multiple transmitters and receivers even if there is some crosstalk between the two channels. In certain cases, when there is nonzero correlated excess noise in the two received signals, we can even surpass the multiplexing gain.
\end{abstract}

\maketitle
\ifpdf
    \graphicspath{{Chapter1/Chapter1Figs/PNG/}{Chapter1/Chapter1/PDF/}{Chapter1/Chapter1Figs/}}
\else
    \graphicspath{{Chapter1/Chapter1Figs/EPS/}{Chapter1/Chapter1/}}
\fi

\section{Introduction}
Quantum key distribution (QKD) enables two remote parties to securely exchange a secret key in the presence of potential eavesdroppers \cite{Pirandola2020AdvancesCryptography}.
Point-to-point QKD links are, however, limited, in terms of secret key rate (SKR) versus distance, to fundamental bounds that depend on the transmissivity of the channel  \cite{Pirandola2017FundamentalCommunications} and the noise in the system. To improve the total SKR, we can use multiplexing techniques to create multiple parallel channels for key exchange \cite{Kumar2019ContinuousDetectors}. Ideally these channels need to be independent, in which case the SKR would increase linearly with the number of channels. In some scenarios, however, this may not be easily feasible. For example, in satellite-based QKD \cite{Hosseinidehaj2019Satellite-basedOutlook} with spatial multiplexing, beam scattering and the scintillation effects in the atmospheric part of the link may result in crosstalk between parallel channels \cite{Kodheli2021SatelliteChallenges}. The same situation may happen in QKD over multi-core fiber \cite{Plews2016QuantumFiber}. In such cases, the key rate per single mode channel may considerably drop because of the crosstalk noise generated by other channels. The above scenarios resemble the situation we face in the well-studied multiple-input multiple-output (MIMO) channels in wireless communications. In this paper, we investigate how MIMO techniques can help QKD systems in harsh channel conditions.

Satellite-to-ground links are often highly lossy for QKD purposes. The geometric loss in free space, augmented by beam wandering and fading effects in the atmospheric part of the link, as well as the limited size of antennas, lead to considerable amount of loss. For example, the total channel loss observed in recent experiments done by the Chinese satellite, Micius, is roughly 30-40~dB \cite{Liao2017Satellite-to-groundDistribution}. This can go up to 85~dB for geostationary satellites \cite{Gunthner2017Quantum-limitedSatelliteb}. This has resulted in focusing mainly on discrete-variable (DV) protocols for satellite-based QKD \cite{Bedington2017ProgressDistribution}, and to discount the continuous-variable (CV) options, which often do not perform well when channel loss is high \cite{Huang2016Long-distanceNoiseb}. 

CV-QKD, however, offers some advantages over DV-QKD if its similarity to coherent communications systems is properly exploited. This could allow us to use well-known techniques in wireless communications to improve system performance. In particular, the combination of CV-QKD with MIMO could be interesting because we can recover phase and amplitude information that are critical to the operation of CV systems. 

In this work, we examine the performance of CV-QKD systems in MIMO settings. 
For simplicity, we consider the $2\times2$ scenario, and model the attack by generalising the ideas in the entangling cloner attack \cite{Navascues2005SecurityDistribution}.
We work out the SKR, and its dependence on relevant parameters, and show how such parameters can be estimated from the corresponding covariance matrix. Our results show that not only the SKR improves in this multi-antenna setting but also the system could become more resilient to loss. 

The rest of the paper is organized as follows. In Sec.~\ref{sec:ProbDes}, we describe the setting of interest in more detail, followed by our security analysis in Sec.~\ref{sec:SecAnal}. We present and discuss our numerical results in Sec.~\ref{sec:NumRes} and conclude the paper in Sec.~\ref{sec:Conc}.

\section{Problem Description}
\label{sec:ProbDes}
\begin{figure}[htbp]
\begin{center}
  \mbox{
     {\includegraphics[width=\linewidth, angle=-0]{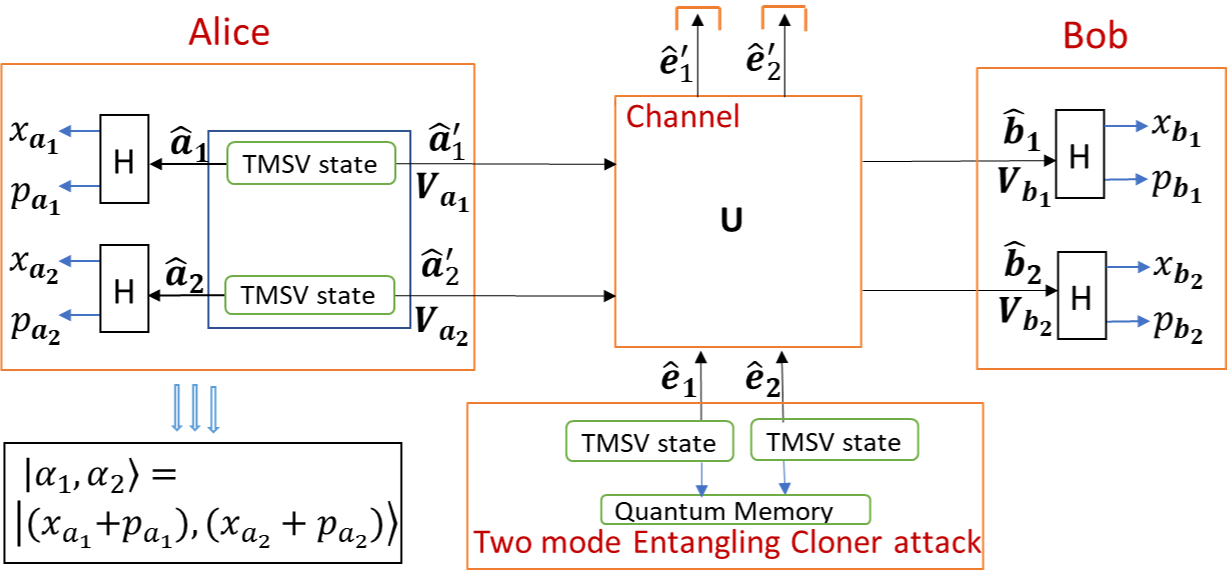}}}
   \caption{Entanglement based picture of the two-mode CV QKD protocol. }
   \label{fig: Two-mode CV QKD}
\end{center}
\end{figure}

In this work, we consider a CV-QKD system over a $2 \times 2$ MIMO channel with two transmitters and two receivers; see Fig.~\ref{fig: Two-mode CV QKD}. Each transmitter independently uses Gaussian encoding to exchange data with its intended receiver. The channel connecting transmitters and receivers could, however, cause interference such that the received signal on each receiver may contain components from both transmitters. For example, in the satellite-based CV QKD with two transmitting antennas, if the corresponding ground stations are in the vicinity of each other, they may each capture part of the signal intended for the other receiver. The additional fading and phase distortions that may also happen in the atmospheric part of the link can further adversely affect system performance. For instance, even if we use only one transmitting antenna, but two nearby ground stations to improve our collection efficiency, the signal received on each telescope may have different phase and amplitude. A simple summation of the two received signals will not then necessarily provide good correlation with the transmitted signal. Similarly, in the $2 \times 2$ case, treating the received signals independently as in two multiplexed systems could see substantial decrease in the SKR due to the crosstalk noise coming from the other channel. In both cases, proper MIMO based postprocessing is needed to extract key information from the received signals. Our objective is to find the SKR achievable in the MIMO setting when such considerations are accounted for.

Here is the detailed description of the protocol used in our MIMO setting, which is based on the CV QKD protocol with Gaussian modulated coherent states and heterodyne detection \cite{Weedbrook2004QuantumSwitching}. There are three parts in our protocol as follows: 

\noindent\textbf{(a) Quantum communication:}  
In this part of the protocol, quantum states are sent over the quantum channel by Alice, the transmitter node, to Bob, the receiver node, to be used for secret key generation. 
Here, Alice prepares two coherent states, $|x_{a_{1}}+i p_{a_{1}}\rangle$ and $|x_{a_{2}}+i p_{a_{2}}\rangle$, where $a_1 = \{x_{a_{1}},p_{a_{1}}\}$ and $a_2 = \{x_{a_{2}},p_{a_{2}}\}$, respectively, represent two sets of i.i.d Gaussian random variables with $0$ mean and variances $V_{a_{1}}$ and $V_{a_{2}}$. Upon receiving the signals, Bob performs a heterodyne measurement on them to, respectively, obtain $b_1 = \{x_{b_{1}},p_{b_{1}}\}$ and $b_2 = \{x_{b_{2}},p_{b_{2}}\}$. This part will be repeated many times to come up, in the asymptotic case, with an infinitely large set of data points.

\noindent\textbf{(b) MIMO processing:} 
In this step, Alice and Bob choose how they wish to process the two sets of data that they have shared via part (a). Here, we consider two cases:
\label{section: Post-processing scenarios}

 \textbf{Case 1, Selection diversity:} Alice and Bob, consider four sets of data $\{(a_1,b_1)\}$,  $\{(a_1,b_2)\}$, $\{(a_2,b_1)\}$ or $\{(a_2,b_2)\}$ to extract the secret key from. The option that offers the highest SKR, let's denote it by $\{(a,b)\}$, is chosen for key exchange. This is akin to selection diversity techniques for receiver spatial diversity in wireless communications.
 
\textbf{Case 2, Full MIMO:} Alice and Bob consider the full set of input-output data, $\{(a_1,a_2),(b_1,b_2)\}$, to obtain secret keys. This step should provide us with a recipe for how to process the above set of data to come up with a correlated set of continuous data points, $\{(a,b)\}$, that can be used by Alice and Bob for key extraction. In our work, we focus on the maximum SKR that can, in principle, be achieved using the ideal MIMO processing.

\noindent\textbf{(c) Classical post-processing:} Once the MIMO processing approach is chosen, discretization, reverse reconciliation and privacy amplification steps are performed on $\{(a,b)\}$ as for the single-mode CV QKD \cite{Weedbrook2004QuantumSwitching}, to obtain a secret key.

In our security analysis, the equivalent entanglement based (EB) picture is considered as shown in Fig.~\ref{fig: Two-mode CV QKD}. In the EB scenario, Alice prepares two two-mode squeezed vacuum (TMSV) states. She measures one mode of each TMSV state using heterodyne detection to project the other mode into coherent states, which she sends to Bob over the quantum channel. The covariance matrix for the initial state of optical modes $A$ and $A'$, respectively, represented, in Fig.~\ref{fig: Two-mode CV QKD}, by annihilation operators $(\hat a_1, \hat a_2)$ and $(\hat a'_1, \hat a'_2)$, is given by
\begin{equation}
\begin{aligned}
\bm{\gamma_{AA'}} = \begin{pmatrix}
V_{a_{1}}\mathds{1}_{2} & k\sigma_{z} \\
k\sigma_{z} & V_{a_{1}} \mathds{1}_{2} \end{pmatrix}\oplus \begin{pmatrix}V_{a_{2}} \mathds{1}_{2}& l\sigma_{z} \\
 l\sigma_{z}& V_{a_{2}} \mathds{1}_{2} 
\end{pmatrix},
\end{aligned}
\label{eq: initial cov mat}
\end{equation}
where $k=\sqrt{V_{a_{1}}^2-1}$ and $l=\sqrt{V_{a_{2}}^2-1}$. Here, $\oplus$ refers to the direct sum, $\mathds{1}_{n}$ is the identity operator of dimension $n$, and $\sigma_{z}$ is Pauli operator $Z$. Note that the first block in $\bm{\gamma_{AA'}}$ corresponds to $(\hat a_1, \hat a'_1)$, whereas the second to $(\hat a_2, \hat a'_2)$.

\section{Security Analysis}
\label{sec:SecAnal}
The secret key rate for a CV-QKD system with reverse reconciliation, in the asymptotic limit where infinitely many signals are exchanged, is given by \cite{Weedbrook2004QuantumSwitching} 
\begin{equation}
    K = \max\{0,\beta \textit{I}(a:b)-\chi_{\small{BE}}\},
\end{equation}
where $I(a:b)$ is the mutual information between Alice's and Bob's data after MIMO processing, $\chi_{BE}$ is Holevo bound on Eve's accessible information on Bob's measurements and $\beta \leq 1$ is the reconciliation efficiency. In a real experiment, $I(a:b)$ can statistically be found using the data exchanged between Alice and Bob. The key problem is then to upper bound the amount of information that has leaked to Eve, i.e., $\chi_{BE}$. This gives us a lower bound on the key rate, which will specify the amount of privacy amplification needed in the protocol. In order to upper bound $\chi_{BE}$, one can use the optimality of Gaussian attacks, which will provide us with a recipe to bound $\chi_{BE}$ using the measured elements of the covariance matrix between Alice and Bob. 

To get an insight into how the key rate behaves, here we consider a special case, which resembles an extension of the entangling cloner attack \cite{Navascues2005SecurityDistribution}. Below we will explain this attack for which we calculate the corresponding covariance matrix and the SKR that can be achieved in Cases 1 and 2 explained in Sec.~\ref{sec:ProbDes}.

\subsection{Eavesdropping attack}
\label{sec:EveAttack}
Here, we take a multimode extension of the entangling cloner attack \cite{Navascues2005SecurityDistribution} as a possible candidate for the optimal attack by Eve for Gaussian encoding. In general, the maximum number of modes required to purify Alice and Bob's modes are less than or equal to twice the number of Alice's modes \cite{Caruso2008Multi-modeChannels}. As a result, for our $2\times2$ MIMO channel, it is sufficient to consider four additional environment/Eve's inputs, for the optimal unitary dilation of the two-mode channel. For this attack, we assume that Eve uses two TMSV states, with variances $V_{e_{1}}$ and $V_{e_{2}}$, as the initial state in the EB picture of Fig.~\ref{fig: Two-mode CV QKD}. One leg of each TMSV state interacts with Alice's signal via a unitary operator $\bm U$, while the other can be stored in quantum memories. This unitary operation effectively replaces the beam splitter in the single-mode case. Due to extremality of Gaussian states \cite{Wolf2006ExtremalityStates} and optimality of Gaussian attacks \cite{Navascues2006OptimalityCryptography ,Garcia-Patron2006UnconditionalDistribution}, and the condition that Eve's state purifies the global state, such a Gaussian attack by Eve maximizes her accessible information.


In Fig.~\ref{fig: Two-mode CV QKD}, the unitary channel transformation matrix $\bm{U}$ maps input annihilation operators to output ones as follows:
\begin{equation}
\begin{aligned}
    \begin{pmatrix}
    \hat b_{1} \\ \hat b_{2}\\\hat e'_{1} \\ \hat e'_{2}
    \end{pmatrix} =\left(
\begin{array}{cc|cc}
    u_{11} & u_{12} & u_{13} & u_{14}\\
    u_{21} & u_{22} & u_{23} & u_{24}\\ \hline
    u_{31} & u_{32} & u_{33} & u_{34}\\
    u_{41} & u_{42} & u_{43} & u_{44}
    \end{array}\right)
    \begin{pmatrix}
    \hat a'_{1} \\ \hat a'_{2} \\ \hat e_{1} \\ \hat e_{2}
    \end{pmatrix}
    \end{aligned},
    \label{eq: annihilation transformations}
\end{equation}
where relevant operators are all specified in Fig.~\ref{fig: Two-mode CV QKD}. We refer to the upper left block in $\bm U$ as the $\bm{H}$-matrix, which effectively replicates the MIMO channel in the classical case.

\subsection{Covariance matrix calculations}
\label{sec:CovMat}
An alternative way of representing the input-output relationship in \eqref{eq: annihilation transformations} is to use the symplectic transformation $\bm{\hat r_{f}}=\bm{S_{fi}}\bm{\hat r_{i}}$, where $\bm{\hat r_{i}} = [\bm{\hat r}_{a'_{1}},\bm{\hat r}_{a'_{2}},\bm{\hat r}_{e_{1}},\bm{\hat r}_{e_{2}}]^{T}$ is the input quadrature operator vector and $\bm{\hat r_{f}} = [\bm{\hat r}_{b_{1}},\bm{\hat r}_{b_{2}},\bm{\hat r}_{e'_{1}},\bm{\hat r}_{e'_{2}}]^{T}$ is the output quadrature operator vector, with $\bm{\hat r}_{z}=[{\hat x_{z}},{\hat p_{z}}]^{T}$, for any annihilation operator $\hat z$, being the vector of the corresponding canonical operators. Here, the symplectic orthogonal matrix $\bm{S_{fi}}$ is given by:
\begin{equation}
    \bm{S_{fi}}
    =\left(
\begin{array}{cc|cc}
  \bm{S}_{11} &   \bm{S}_{12} &   \bm{S}_{13} &   \bm{S}_{14}\\
  \bm{S}_{21} &   \bm{S}_{22} &  \bm{S}_{23} &   \bm{S}_{24}\\ \hline
  \bm{S}_{31} &   \bm{S}_{32} &   \bm{S}_{33} &   \bm{S}_{34}\\
  \bm{S}_{41} & \bm{S}_{42} &  \bm{S}_{43} &   \bm{S}_{44}
\end{array}
\right), 
 \\
 \label{eq: s1}
\end{equation}
where
\begin{equation}
    \bm{S}_{mn} = \left(\begin{array}{cc}
\Re\{u_{mn}\} &  -\Im\{u_{mn}\}\\
\Im\{u_{mn}\} & \Re\{u_{mn}\}
 \end{array}\right), \quad \mbox{$m,n= 1, \ldots, 4$}.
\end{equation}
Using the transformation in \eqref{eq: s1}, we can now obtain the corresponding covariance matrix between Bob's modes, $\bm B$, and the two modes that Alice retains in the EB picture, $\bm A$, as follows:
\begin{equation}
\bm{\gamma_{AB}} = \trace_{E}[\bm{S} (\bm{\gamma_{AA'}}\oplus\bm{\gamma_{E}}) \bm{S}^{T}],
\label{eq: Stinespring dilation cov mat}
\end{equation}
where $\bm{S} = \mathds{1}_{4}\oplus \bm{S_{fi}}$, to which necessary re-arrangements must be applied to match the coordinates of other matrices in \eqref{eq: Stinespring dilation cov mat}, $\bm{\gamma_{AA'}}$ is the initial covariance matrix for Alice as given in \eqref{eq: initial cov mat},  $\bm{\gamma_{E}}= V_{e_{1}} \mathds{1}_{2}\oplus V_{e_{2}} \mathds{1}_{2}$, and $\trace_E\left[*\right]$ is obtained by excluding the rows and columns corresponding to Eve's modes.

Using \eqref{eq: annihilation transformations}--\eqref{eq: Stinespring dilation cov mat}, the covariance matrix for modes $\bm A$ and $\bm B$ is given by:
\begin{equation}
    \bm{\gamma_{AB}} =\begin{aligned} \begin{blockarray}{ccccc}
\text{modes:} &a_1&b_1&a_2&b_2\\
\begin{block}{c(cccc)} 
a_1&\bm{\gamma_{A_{1}}} & \bm{\gamma_{A_{1}B_{1}}} & \bm{\gamma_{A_{1}A_{2}}} & \bm{\gamma_{A_{1}B_{2}}} \\
b_1&\bm{\gamma_{A_{1}B_{1}}} & \bm{\gamma_{B_{1}}} & \bm{\gamma_{A_{2}B_{1}}} & \bm{\gamma_{B_{1}B_{2}}} \\
a_2&\bm{\gamma_{A_{1}A_{2}}} & \bm{\gamma_{A_{2}B_{1}}} & \bm{\gamma_{A_{2}}} & \bm{\gamma_{A_{2}B_{2}}} \\
a_2&\bm{\gamma_{A_{1}B_{2}}} & \bm{\gamma_{B_{1}B_{2}}^{T}} & \bm{\gamma_{A_{2}B_{2}}} & \bm{\gamma_{B_{2}}} \\
\end{block}
\end{blockarray}
\end{aligned}
\label{eq: cov mat for genric symplectic transform}
\end{equation}
where the covariance matrix elements are given by:
\begin{eqnarray}
&\bm{\gamma_{A_{1}}}=V_{a_1}\mathds{1}_{2}, \bm{\gamma_{A_{2}}}=V_{a_2}\mathds{1}_{2},& \nonumber\\
&\bm{\gamma_{A_{1}B_{1}}}=k \bm{F}(u_{11}),
\bm{\gamma_{A_{1}B_{2}}}=k \bm{F}(u_{21}),& \nonumber\\
&\bm{\gamma_{A_{2}B_{1}}}=l\bm{F}(u_{12}),
\bm{\gamma_{A_{2}B_{2}}}=l\bm{F}(u_{22}),\bm{\gamma_{A_{1}A_{2}}}=0,& 
\label{eq: mimo cov mat elements}
\end{eqnarray}
with 
\begin{equation}
    \bm{F}(u)=\begin{pmatrix}
\Re\{u\} & \Im\{u\} \\
\Im\{u\} & -\Re\{u\}
\end{pmatrix},
\end{equation}
and
\begin{equation}
\bm{\gamma_{B_{1}}}=\delta_1 \mathds{1}_{2} ,
\bm{\gamma_{B_{2}}}=\mu_1 \mathds{1}_{2},
\bm{\gamma_{B_{1}B_{2}}}=\begin{pmatrix}
\nu_1 & \nu_3 \\
-\nu_3 & \nu_1
\end{pmatrix},
\end{equation}
with
\begin{eqnarray}
&\delta_1
=\bm{v_1^\dagger{v_1}}=|u_{11}|^2f_1+|u_{12}|^2f_2+1+\xi_{b_{1}},&\nonumber\\
&\mu_1
=\bm{v_2^\dagger{v_2}}=|u_{21}|^2f_1+|u_{22}|^2f_2+1+\xi_{b_{2}},&\nonumber\\
&\begin{aligned}
\nu_1+i\nu_3
=\bm{ v_1^\dagger v_2}
=u_{11}^*u_{21}f_1+u_{12}^*u_{22}f_2+\xi_{b_{1}b_{2}},\end{aligned}&
\label{eq: PS+BS covmat with thermal Eve}
\end{eqnarray}
where $\xi_{b_{1}}$ and $\xi_{b_{2}}$ are, respectively, the excess noise at Bob's first and second receiver, $\xi_{b_{1}b_{2}}$ represents a cross-correlation excess noise term between the two receivers, and
\begin{eqnarray}
&f_{1}=(V_{a_{1}}-1),\quad f_2=(V_{a_{2}}-1),&\nonumber\\
&\bm{{v_1}}=[\sqrt{V_{e_1}}u_{13} \quad \sqrt{V_{e_2}}u_{14} \quad \sqrt{V_{a_1}}u_{11} \quad \sqrt{V_{a_2}}u_{12} ]^T,&\nonumber\\
&\bm{{v_2}}=[\sqrt{V_{e_1}}u_{23} \quad \sqrt{V_{e_2}}u_{24} \quad \sqrt{V_{a_1}}u_{21} \quad \sqrt{V_{a_2}}u_{22} ]^T.&
\end{eqnarray}

Given that all elements of the covariance matrix can be measured in a real experiment, there are several interesting observations we can make from the covariance matrix elements:
\begin{itemize}
    \item  It is interesting to see that the MIMO channel, $\bm H$, can be obtained directly from the elements presented in \eqref{eq: mimo cov mat elements}. This channel parameter estimation fully relies on the quantum communication part of the protocol, and can therefore be used for bounding the Holevo information term.
    \item Similar to the single-mode CV-QKD case, here, we have defined the excess noise at each receiver as the difference between the noise measured at each receiver and the minimum noise that we expect assuming that there is no Eve present, i.e., in our model, when $V_{e_1} = V_{e_2} = 1$ in shot-noise units (SNU). 
    \item We, however, have a new excess noise observable, $\xi_{b_{1}b_{2}}$, that does not exist in single-mode CV-QKD. This new parameter models the correlated excess noise between the two receivers. It would be interesting to see how this parameter affects system performance. We refer to the cases where $\xi_{b_{1}b_{2}} \neq  0$ as the colored, or correlated, excess noise case. This is in contrast to the typical case in classical MIMO where noise at the receiver is modelled by i.i.d. random variables, and we will see how it can actually offer additional improvement in performance.
\end{itemize}


While the values observed for $\xi_{b_1}$ and $\xi_{b_2}$ can, in principle, be any non-negative real number, the values that $\xi_{b_{1}b_{2}}$ can take should satisfy the following restrictions:
\begin{itemize}
    \item[(1)] All symplectic eigenvalues of $\bm{\gamma_{AB}}$ should be greater than or equal to 1 \cite{Simon1994Quantum-noiseForms, Williamson1936OnSystemsb}.
    \item[(2)] By applying Cauchy-Schwarz inequality 
\begin{equation}
    |\bm{ v_1^\dagger v_2}|^2 \leq {(\bm{ v_1^\dagger v_1}) (\bm{ v_2^\dagger v_2})} 
\end{equation}
to \eqref{eq: PS+BS covmat with thermal Eve}, we obtain the following relation for the permissible region of $\xi_{b_{1}b_{2}}$:
\begin{equation}
    \nu_1^2+\nu_3^2\leq \delta_1 \mu_1.
    \label{eq: cauchy-schwarz}
\end{equation}
\end{itemize}

\subsection{Key rate analysis}
\label{Multimode Key rate analysis}
Based on the channel model given in Sec.~\ref{sec:EveAttack} and the covariance matrix obtained in Sec.~\ref{sec:CovMat}, we can now work out the SKR in the two cases introduced in Sec.~\ref{sec:ProbDes}. 

\textbf{Case 1:} Here, we calculate the key rate for one of the four cases mentioned under selection diversity. The SKR in the other cases can similarly be calculated to the case $\{a_1,b_1\}$ described here. The secret key rate for this case is given by:
\begin{equation}
    K(a_{1},b_{1})=\beta  I({a_{1}}:{b_{1}})-\chi({b_{1}}:E),
    \label{eq: MIMO case1 SKR}
\end{equation}
where $E$ represents all four quantum modes that Eve has access to in Fig.~\ref{fig: Two-mode CV QKD}. In this case, we assume the state shared by Alice and Bob is Gaussian, in which case the mutual information term is given by:
     \begin{equation}
             I(a_{1}:b_{1}) =\frac{1}{2}\log_{2}\left[\frac{{\rm Det}[V_{A_{1}}]{\rm Det}[V_{B_{1}}]}{{\rm Det}[V_{A_{1},B_{1}}]}\right]         ,
    \end{equation} 
where $V_{A_1}$, $V_{B_1}$, and $V_{A_{1}B_{1}}$ are, respectively, the covariance matrix for Alice's mode $A_{1}$, Bob's mode $B_{1}$ and the joint covariance matrix for the modes $A_1$ and $B_1$. These matrices can be found by applying heterodyne measurement on the covariance matrix $\bm{\gamma_{AB}}$, given as
\begin{equation}
    \bm{\gamma_{AB_{het}}}=\frac{\bm{\gamma_{AB}}+\mathds{1}_8}{2}, 
    \label{eq: MI cov mat for genric symplectic transform}
\end{equation}
 and keeping the relevant modes.
 
The Holevo bound on Eve's information over Bob's measurement is given by:
    \begin{equation}
    \begin{aligned}
    \chi({b_{1}}:E)&=S(E)-S(E|{b_{1}})\\
 &=S(A_{1}B_{1}A_{2}B_{2})-S(A_{1}A_{2}B_{2}|{b_{1}}),
 \end{aligned}
 \end{equation} 
where, $S$ represents the von Neumann entropy function. There are standard techniques to calculate this function for Gaussian states, which has been summarised in Appendix \ref{sec:App}. The two entropy terms above can then numerically be found using \eqref{eq: Von Neumann entropy Gaussian st}-\eqref{eq: conditional heterodyne detection}. 
    
Note that, in practice, using selection diversity may make sense if one channel is considerably better than the other one, as otherwise part of the signal sent by the second transmitter could enter as noise on the first receiver. Alternatively, one may consider the multiplexing option in which case the total key rate is given by $K_{total}=K(a_{1}:b_{1})+K(a_{2}:b_{2})$ or  $K_{total}=K(a_{1}:b_{2})+K(a_{2}:b_{1})$ depending on channel conditions.

\textbf{Case 2:} Using the full information available to the users, the SKR is given by:
    \begin{equation}
        K(a_{1},a_{2}:b_{1},b_{2})=\beta  I({a_{1},a_{2}:b_{1},b_{2}})-\chi({b_{1},b_{2}}:E).
        \label{eq:keyrateCase2}
    \end{equation}
Again, under Gaussian assumption for the joint state of Alice and Bob, the mutual information for this case is given by:
     \begin{equation}
      I(a_{1},a_{2}:b_{1},b_{2})=
     \frac{1}{2}\log_{2}\left[\frac{{\rm Det}[V_{A_{1},A_{2}}]{\rm Det}[V_{B_{1},B_{2}}]}{{\rm Det}[V_{A_{1},A_{2},B_{1},B_{2}}]}\right].\\
    \end{equation}
All relevant covariance matrices in the above equation can be found using the covariance matrix  $\bm{\gamma_{AB_{het}}}$, given in \eqref{eq: MI cov mat for genric symplectic transform}. The Holevo bound is also given by:
    \begin{equation}
    \begin{aligned}
    \chi({b_{1},b_{2}}:E)&=S(E)-S(E|{b_{1},b_{2}})\\
    &=S(A_{1}B_{1}A_{2}B_{2})-S(A_{1}A_{2}|{b_{1},b_{2}}),
    \end{aligned}
    \end{equation}
where, again, by finding the relevant symplectic eigenvalues, we can use \eqref{eq: Von Neumann entropy Gaussian st}-\eqref{eq: conditional heterodyne detection} to numerically calculate the entropy terms in the above equation. Note that the key rate in \eqref{eq:keyrateCase2} is the maximum that can be obtained in the MIMO setting considered conditioned on using the optimal MIMO postprocessing applicable to the measured data points. 
\section{Key Rate Simulation: Results and Discussion}
\label{sec:NumRes}
In this section, we compare the key rate for the $2\times2$ MIMO CV-QKD with that of single-mode CV-QKD.
We consider a MIMO channel in which the line of sight link carries the same power as the crosstalk one given by the following channel state matrix:
\begin{equation}
    \bm{H}=\sqrt{\frac{T}{2}}\begin{pmatrix}
1 & i \\
i & 1
\end{pmatrix},
\label{eq: H-matrix}
\end{equation}
where $T$ models the channel transmissivity. We assume the excess noise $(\xi_{b_1},\xi_{b_2})$ at both receivers to be constant and not dependent on the channel loss. For our simulation, we have taken $\xi_{b_1}=\xi_{b_2}=0.001$ SNU
, and the reconciliation efficiency $\beta$ is $0.95$. We optimize the SKR over $V_{a_1}$ and $V_{a_2}$ under the constraint that the maximum total power is $2\times4.7$~SNU. This enables the comparison between the single-mode and two-mode MIMO cases, as 4.7~SNU is the optimum power transmitted in the single-mode case at high losses for our chosen parameter values. It turns out that the optimum power allocation in the MIMO setup is the same as equal power allocation on both transmitted modes. 

\begin{center}
\begin{figure}[htbp]
     \includegraphics[width=\linewidth, angle=-0]{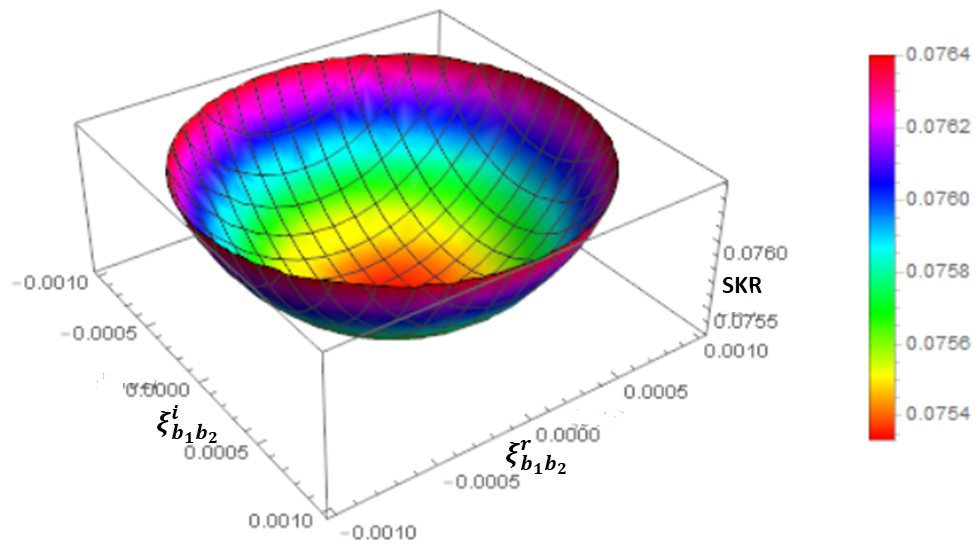}
    \caption{The MIMO secret key rate (SKR) versus real, $\xi_{b_1b_2}^r$, and imaginary, $\xi_{b_1b_2}^i$, parts of $\xi_{b_{1}b_{2}}$, in its permissible region, at $T=0.1$.}
   \label{fig: MIMO SKR(2)- permissible reg Plots}
\end{figure}\end{center}
Let us begin by looking into the effect of the correlated noise parameter, $\xi_{b_1b_2}$, on the key rate. Fig. \ref{fig: MIMO SKR(2)- permissible reg Plots} shows the SKR vs $\xi_{b_1b_2}$. It can be seen that, for the particular channel considered here, 
the SKR is minimum when $\xi_{b_{1}b_{2}} = 0$. SKR is maximum when $\xi_{b_{1}b_{2}}$ reaches the values on the circular boundary, which satisfy the condition \eqref{eq: cauchy-schwarz}.
 This suggests that the worst case scenario for Alice and Bob, for this channel, happens when the attack by Eve leaves no excess correlation in the noise terms. In the rest of the paper, unless otherwise noted, we therefore assume $\xi_{b_1b_2}$ = 0, which corresponds to the minimum SKR for the given channel parameters. For the colored noise environment, we obtain the SKR plots, optimized over the input modulation variances {$V_{a_1}, V_{a_2}$}, by taking $\xi_{b_{1}b_{2}}=0.0006+i 0.00079$, which lies on the circular boundary of values that lead to the maximum SKR. 

\begin{center}
\begin{figure}[htbp]
     \mbox{\includegraphics[width=3.35in, angle=-0]{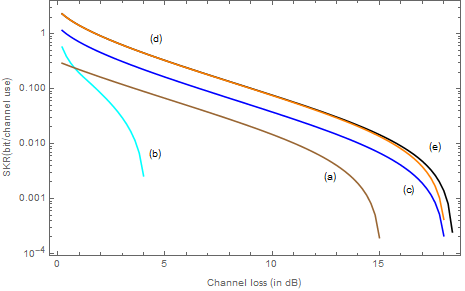}}
    \caption{The secret key rate (SKR) versus channel loss, $1/T$, for different scenarios. Labels are defined in the text.}
   \label{fig: Two-mode CV QKD key-rate Plots}
\end{figure}\end{center}

In Fig.~\ref{fig: Two-mode CV QKD key-rate Plots}, we compare the optimized SKR vs channel loss parameter, $1/T$, in different scenarios of interest. These cases include (labels on the graph correspond to the labels below in the text):

(a) The SKR (brown curve) when one of the transmitters is in use, while the other mode is off. In this case, there is no interference/crosstalk from the other channel, but the channel transmissivity is equal to $H_{11}=\sqrt{T/2}$, which accounts for the fact that the channel suffers from scattering issues. 


(b) The total SKR (cyan curve) in the multiplexed scenario in the presence of crosstalk noise, i.e, the sum of  $K(a_1,b_1)$ and $K(a_2,b_2)$, calculated using \eqref{eq: MIMO case1 SKR}, when $V_{a_1}, V_{a_2}$ are varied up to $4.7$~SNU. 
The direct channel transmissivities are $H_{11} = H_{22}=\sqrt{T/2}$, while the overall channel $H$-matrix is the same as that given in \eqref{eq: H-matrix}. Because of taking into account the noise contribution from other channels, this SKR behaves poorly as compared to other cases. 

(c) The SKR (blue curve) when the channel is a single-mode CV-QKD channel with transmissivity $\sqrt{T}$, and in the absence of crosstalk from any other channel. This is the maximum secret key rate that can be obtained in a single-input single-output (SISO) CV QKD channel when there is no scattering and no crosstalk issues in the channel.

(d) The SKR (orange curve) in the $2\times2$ MIMO channel when the noise at the receiver is i.i.d., i.e., $\xi_{b_1b_2}$= 0. The SKR in this case turns out to be equal to the sum of key rates for both channels when the MIMO channel is converted into two independent channels via the singular value decomposition (SVD) of the $H$-matrix\cite{Kundu2021MIMODistribution}. This is twice the SKR for the SISO case (c), as the singular values for the $H$-matrix we have considered are equal to the transmissivity for the single-mode channel. Thus it is equivalent to the multiplexing gain of factor 2 as compared to the SISO secret key rate of case (c). This is considerably better than case (a), which also accounts for additional scattering issues in the SISO case for the particualr channel matrix considered here.

(e) The SKR (black curve) in the MIMO channel in the presence of crosstalk and correlated/colored noise, i.e., when $\xi_{b_1b_2}$ takes an optimal non-zero value. This offers a key rate slightly more than the i.i.d. noise scenario of case (d), as per the results in Fig.~\ref{fig: MIMO SKR(2)- permissible reg Plots}. Thus the gain factor is more than 2 in this case, as compared to the SISO SKR of case (c).

From our numerical simulations, we observe that in the scenarios with some crosstalk between the employed channels, MIMO processing can significantly improve the performance as compared to multiplexing techniques, where post-processing is done on two separate single-mode CV-QKD systems \cite{Kumar2019ContinuousDetectors}. In the latter case, the crosstalk signal from one channel is treated as the excess noise in the other system which can considerably hamper its performance. When we do the MIMO post-processing, a new set of observables, including the crosstalk coefficients and correlations that may be observed between the two received signals, can help us with key extraction. 
 

\section{Conclusions}
\label{sec:Conc}
We proposed CV QKD over a $2 \times 2$ MIMO setting. Setups like this could become relevant in satellite-based QKD. We performed the security analysis for a Gaussian encoded protocol and investigated the SKR versus loss behaviour for a particular MIMO channel with equal distribution of power between the line-of-sight and crosstalk links. It turned out that by using the full MIMO processing power we could mostly remove the crosstalk effect and recover the multiplexing gain of two. Even more, when there was correlated excess noise between the two receivers, we could obtain a slightly higher amount of secret keys. While this work needs to be extended to account for channels of higher dimensions, among other things, our results suggest that MIMO techniques can offer a promising approach to improving SKR performance in scenarios where phase and amplitude distortions are inevitable.

\appendix
\section{Gaussian states}
\label{sec:App}
Here, we summarize the key techniques for calculating the von Neumann entropy of a Gaussian state.

\textbf{1) Entropy of Gaussian states:}
The von Neumann entropy of an $N$-mode Gaussian state $\rho$, with a covariance matrix $\bm{V}_\rho$, is given by:
\begin{equation}
    S({\rho})=\sum_{n=1}^{N} g(\lambda_{n}),
    \label{eq: Von Neumann entropy Gaussian st}
\end{equation}
where
\begin{equation}
\label{eq: thermal entropy}
    g(\lambda_{n})=\frac{\lambda_{n}+1}{2}\log_{2}\left[\frac{\lambda_{n}+1}{2}\right]-\frac{\lambda_{n}-1}{2}\log_{2}\left[\frac{\lambda_{n}-1}{2}\right],
\end{equation}
and $\lambda_{n}$s are the symplectic eigenvalues of $\bm{V}_\rho$. These eigenvalues can be found by finding the eigenvalues of the matrix $|i\bm{\Omega V}_{\rho}|$, i.e., the positive eigenvalues of the matrix $i\bm{\Omega V}_{\rho}$, where $\bm{\Omega}= \bigoplus_{i=1}^n \begin{pmatrix}
0 & 1 \\
-1 & 0
\end{pmatrix}$.

\textbf{2) Conditional covariance matrix:}
Given a heterodyne measurement on a single mode $B$ of an $N$-mode Gaussian system $AB$, with covariance matrix $\gamma_{AB}$, the conditional covariance matrix denoted by $\gamma_{A|B}$ is given by \cite{Serafini2017QuantumMethods}:
\begin{equation}
    \gamma_{A|B}=\gamma_{A}-\gamma_{AB}(\gamma_{B}+\mathds{1}_{2})^{-1}\gamma_{AB}^{T},    
\label{eq: conditional heterodyne detection}    
\end{equation}
where $\gamma_{A}$ ($\gamma_{B}$) is the covariance matrix for system $A$ ($B$).

\section*{Acknowledgement}
M.R. is grateful to Masoud Ghalaii for fruitful discussions.  All data generated in this paper can be reproduced by the provided methodology and equations.

\bibliographystyle{ieeetr}
\renewcommand{\bibname}{References} 
\bibliography{references1}
\end{document}